\documentclass[12pt,a4paper]{article}

\usepackage{amsmath}
\usepackage{graphicx}

\makeatletter
\def\Ginclude@eps#1{%
 \message{<#1>}%
  \bgroup
  \def\@tempa{!}%
  \dimen@\Gin@req@width
  \dimen@ii.1bp%
  \divide\dimen@\dimen@ii
  \@tempdima\Gin@req@height
  \divide\@tempdima\dimen@ii
    \includegraphics{#1}%
  \egroup}
\makeatother

\begin{document}

\title{The geometrical origin of the Doppler factor in the Li\'{e}nard-Wiechert potentials}
\author{C\u alin Galeriu}

\maketitle

\section*{Abstract}

We present an in depth analysis and a new derivation of the Doppler factor in the Li\'{e}nard-Wiechert potentials,
based on geometrical considerations in Minkowski space.
We argue that, contrary to a common assumption, 
the methods used for deriving the Doppler factor in the case of an electrically
charged extended particle are not applicable in the case of a point particle.
A geometrical interpretation of the Doppler factor is nonetheless found in the later case,  
based on the electromagnetic interaction model of Fokker.
In this model the interaction takes place between
infinitesimal worldline segments with end points connected by light signals,
just like the points where light pulses are emitted and detected in general. 
This analogy reveals that the relativistic Doppler effect
is the missing link between 
the classical Doppler effect and
the Doppler factor in the Li\'{e}nard-Wiechert potentials.

\section{Introduction}

The Li\'{e}nard-Wiechert (LW) potentials of a moving point charge \cite{lienard,wiechert}
can be derived from the potentials of a stationary electric charge. The transition from the static to the mobile case can happen in different ways, depending on how we look at the electrically charged particle. The particle can be a material point with an electric charge $Q$, or it can extend in space over a small volume, which in the limit collapses into a point. In the later case the particle is described by an electric charge density $\rho(\overrightarrow{r})$ whose integral in space gives the total electric charge $Q$.
\begin{equation}
\int \rho(\overrightarrow{r}) \ dx \ dy \ dz = Q.
\label{eq:1}
\end{equation}

For an electrically charged point particle at rest at point $C$ with position 
$\overrightarrow{r_C} = (x_C, y_C, z_C)$,
the electric (scalar) potential (in Gaussian units) at point $P$ with position
$\overrightarrow{r_P} = (x_P, y_P, z_P)$ 
is $\phi(\overrightarrow{r_P}) = Q / R$, where 
$\overrightarrow{R} = \overrightarrow{r_P} - \overrightarrow{r_C} = (x_P - x_C, y_P - y_C, z_P - z_C)$.
In the electrostatic case the magnetic (vector) potential is $\overrightarrow{A}(\overrightarrow{r_P}) = 0$, and the electromagnetic four-potential at $P$ is
\begin{equation}
\mathbf{\Phi}_P = (\overrightarrow{A}, i \phi) = (0, 0, 0, i \frac{Q}{R}).
\label{eq:2}
\end{equation}

In Minkowski space the field point $P$ has a position four-vector $\mathbf{X}_P = (x_P, y_P, z_P, i c t_P)$, while the stationary source charge at $C$ has a retarded position four-vector $\mathbf{X}_C = (x_C, y_C, z_C, i c t_C)$ and a four-velocity $\mathbf{V}_C = (0, 0, 0, i c)$. Since the two points are connected by a light signal, $(\mathbf{X}_P - \mathbf{X}_C) \cdot (\mathbf{X}_P - \mathbf{X}_C) = 0$, and therefore $R = c (t_P - t_C)$. It follows that  $(\mathbf{X}_P - \mathbf{X}_C) \cdot \mathbf{V}_C = - c R$. The electromagnetic four-potential (\ref{eq:2}) at $P$ becomes
\begin{equation}
\mathbf{\Phi}_P = - \frac{Q\,\mathbf{V}_C}{(\mathbf{X}_P - \mathbf{X}_C) \cdot \mathbf{V}_C}.
\label{eq:3}
\end{equation}

For an electrically charged point particle in motion, with a retarded velocity $\overrightarrow{v} = (v_x, v_y, v_z)$ and a retarded four-velocity 
$\mathbf{V}_C = (\gamma v_x, \gamma v_y, \gamma v_z, i \gamma c)$, 
where $\gamma = (1 - v^2/c^2)^{-1/2}$,
the electromagnetic
four-potential (\ref{eq:3}) at $P$ becomes
\begin{equation}
\mathbf{\Phi}_P = \frac{Q}{R} \frac{(v_x / c, v_y / c, v_z / c, i)}{1 - \overrightarrow{R} \cdot \overrightarrow{v} / (R c)}.
\label{eq:4}
\end{equation}
The denominator of the Doppler factor may also be written 
as $1 - \hat{R} \cdot \overrightarrow{v} / c$ \cite{Griffiths4ed} 
(where $\hat{R} = \overrightarrow{R} / R$ is a radial unit vector), 
or as $1 - v_r / c$ \cite{ORahilly} 
(where $v_r =  \overrightarrow{v} \cdot \hat{R}$ is the radial component of the retarded velocity, directed toward the field point), 
or as $1 - \beta \cos(\theta)$ \cite{Aguirregabiria}
(where $\beta = v / c$ and $\theta$ is the angle between $\overrightarrow{v}$ and $\overrightarrow{R}$).

The electrostatic potential $\phi(\overrightarrow{r_P})$ produced at $P$ by a continous electric charge distribution at rest is
\begin{equation}
\phi(\overrightarrow{r_P}) = \int \frac{\rho(\overrightarrow{r})}{R} \ dx \ dy \ dz,
\label{eq:5}
\end{equation}
where $\overrightarrow{R} = \overrightarrow{r_P} - \overrightarrow{r}$. When $\rho(\overrightarrow{r}) \neq 0$
only in a very small neighborhood centered on point $C$, $R$ gets out of the integral and, because of (\ref{eq:1}),
we get the same electromagnetic four-potential (\ref{eq:2}).

When the continuous electric charge distribution is in motion, the retardation condition changes (\ref{eq:5}) into
\begin{equation}
\phi(\overrightarrow{r_P}, t_P) = \int \frac{\rho(\overrightarrow{r_{ret}}, t_{ret})}{R} \ dx \ dy \ dz,
\label{eq:6}
\end{equation}
where $\overrightarrow{R} = \overrightarrow{r_P} - \overrightarrow{r_{ret}}$ and $t_{ret} = t_P - R / c$. 
When $\rho(\overrightarrow{r_{ret}}, t_{ret}) \neq 0$
only in a very small neighborhood centered on point $C$, $R$ gets out of the integral and we get 
the same electric potential as in (\ref{eq:4}), provided that
\begin{equation}
\int \rho(\overrightarrow{r_{ret}}, t_{ret}) \ dx \ dy \ dz = \frac{Q}{1 - v_r / c}.
\label{eq:7}
\end{equation}

When comparing (\ref{eq:7}) with (\ref{eq:1}), \lq\lq most undergraduate students have major difficulties understanding where the additional factor [...] comes from.\rq\rq \cite{Aguirregabiria} The correct explanation is that the integral in (\ref{eq:7}) 
\lq\lq does not represent the total 
particle charge because the charge density in the integrand is taken at different retarded times\rq\rq.
We have to recognize the fact that the integral in (\ref{eq:7}) \lq\lq is not a simultaneous integral.\rq\rq \cite{Aguirregabiria}

No wonder this step in the derivation of the LW potentials is counterintuitive, 
since in Newtonian physics a volume integration is always done at the same time. In relativistic physics, due to
the relativity of simultaneity, things change a little. 
Often a 3D volume integration is done at the
same time, but only in a specific reference frame. This does not apply here, because there is no reference frame in which the retarded electric charges on the lightcone are simultaneous. There is no such a reference frame because we cannot have a Lorentz boost with the speed of light.  
For a Lorentz boost with the speed of light, the spatial volume element would decrease to zero, while the electric charge density would increase to infinity. Agurregabiria {\it et al.} \cite{Aguirregabiria} mention that 
\lq\lq one can change variables to have a simultaneous integral. 
This gives the factor $( 1 - \beta cos \theta )^{-1}$ as the Jacobian of the transformation, but
the actual computation is a bit cumbersome\rq\rq, 
probably referring to a derivation by O'Rahilly that makes use of curvilinear coordinates. \cite{ORahilly}
Sometimes a 3D volume integration is done on an invariant hypersurface, a 3D subspace of the 4D Minkowski space. This does not apply here either, because the infinitesimal volume element on the lightcone is zero, and as a result the volume element in the integral would be zero.

New concepts are needed, and Aguirregabiria {\it et al.} \cite{Aguirregabiria} describe \lq\lq the effective integration region\rq\rq
as the region where the retarded electric charge density is different from zero, different from
\lq\lq the spatial region occupied by the charge at a single instant of time\rq\rq .
Griffiths \cite{Griffiths4ed} uses an equivalent term, \lq\lq the apparent volume\rq\rq of the moving extended object, 
different from \lq\lq the actual volume\rq\rq of the same object. 

Here we show that a very clear understanding of the concept of effective integration region emerges
when the integral in (\ref{eq:7}) is evaluated in Minkowski space. We use geometrical operations, intersections and projections,
in order to justify the apparition of the Doppler factor in the LW potentials. As stated by Griffiths, 
\lq\lq this is a purely {\it geometrical} effect.\rq\rq \cite{Griffiths4ed} 
A similar geometrical approach (based on intersections) was used by 
Aguirregabiria {\it et al.} \cite{Aguirregabiria} in order to find the retarded shape of a moving sphere.

\section{The conceptual framework}

A first question to ask, when looking at (\ref{eq:7}), is: What is the domain of integration? To what 3D space 
does the $dx \ dy \ dz$ volume element really belong to? The domain of integration is the 3D space simultaneous with the field point $P$ where the LW potentials are calculated. It is the hyperplane of constant time $t = t_P$, a 3D spatial slice through the 4D Minkowski space. 

This fundamental fact is hidden from plain sight by many authors who look at a volume element at the retarded time $t = t_C$. 
Morse and Feshbach \cite{Morse} consider a volume element sandwiched between two spherical boundaries 
drawn at two different retarded times,
and mention that \lq\lq in the integration, the amount of charge $dq$ inside the volume element $dA\ dr_r$ is not 
$\rho\ dA\ dr_r$, as it would be if the charge were not moving, but is $[1 + (u/c) cos \beta] \rho\ dA\ dr_r$\rq\rq, 
as if the electrically charged matter could move in or out of this multi-temporal volume element. 
This construction of the two spherical boundaries closely mirrors that of Li\'{e}nard \cite{lienard} and Wiechert\cite{wiechert},
who compare the volume of this retarded volume element as seen from the stationary reference frame of the observer
with the volume of the same retarded volume element as seen from the moving reference frame of the electrically charged extended particle.
Li\'{e}nard mentions that during the integration in (\ref{eq:7}) the content of this retarded volume element is swept (\lq\lq balay\'{e}\rq\rq) 
by a spherical surface centered on the field point. Panofsky and Phillips \cite{Panofsky}
replace the static picture of the two spherical boundaries drawn at two different retarded times with 
the dynamic picture of one spherical surface collapsing with the speed of light $c$ toward the observation point.
They mention that \lq\lq during the time the information-collecting sphere [...] sweeps over the charge distribution
the charges may move so as to appear more or less dense\rq\rq and then conclude that \lq\lq the retarded potential of an approaching
charge will be larger than that of a receding charge at the same distance from the observer, since the approaching charge
stays longer within the information-collecting sphere\rq\rq. 
The Doppler factor is the result of a multi-temporal electric charge density. 
In the same spirit Feynman writes that \lq\lq there is a correction term which comes about 
because the charge is moving as our integral \lq sweeps over the charge.\rq \ \rq\rq \cite{FeynmanII}
In yet another derivation, again based on a retarded multi-temporal volume element,
Page and Adams \cite{Page} describe how, due to the velocity of the electrically charged matter, the retarded rectangular volume element 
changes shape in order to accommodate the retardation condition on a pair of opposing sides, thus becoming a slanted prism. 
When contemplating all these derivations of the Doppler factor, one may very easily become confused. 

A second question to ask, when looking at (\ref{eq:7}), is: What do we get when we replace the retarded electric charge density in the integral with the instantaneous electric charge density? In this case the integrand is non-zero only inside the volume where the electric charge is present at the actual moment $t = t_P$. This space region is the intersection of the worldtube of the charged particle with the Minkowski hyperplane of constant time $t = t_P$. When the charged particle is at rest, this intersection is a sphere. When the charged particle is in motion, this intersection is a Lorentz ellipsoid. In both cases the total electric charge of the particle is the same. 

A third question to ask, when looking at (\ref{eq:7}), is: How do we calculate this non-simultaneous integral?
At time $t = t_P$, for each point $(x, y, z)$ at the center of a volume element $dx \ dy \ dz$, we need to find out the value of
the retarded electric charge density $\rho(x, y, z, t_{ret})$. 
We have to start at that point, and then go back in time, keeping the same $x, y, z$ coordinates, 
until we meet the retarded lightcone drawn through the field point. 
What is the retarded electric charge density at this place? 
In other words, is the intersection point that we get in this way also a point inside the worldtube of the electrically charged particle? 
In order to answer this question, we realize that we can travel up and down along this path in both ways. 
In Minkowski space, we may start from the $(x, y, z, i c t_P)$ point, go down to the $(x, y, z, i c t_{ret})$ point on the retarded lightcone, 
and then see whether this point belongs to the worldtube of the electrically charged particle.
Or we may start with the intersection of the retarded lightcone with the worldtube of the electrically charged particle, 
and then project this intersection on the Minkowski hyperplane of constant time $t = t_P$. 
We will use the second method in order to find the volume of the region with non-zero retarded electric charge density. 

In conclusion, the effective integration region of Aguirregabiria {\it et al.} \cite{Aguirregabiria},
same as the apparent volume of Griffiths \cite{Griffiths4ed},
is the projection on the field point's 3D space of the intersection of the retarded light cone
with the worldtube of the electrically charged particle. Since the electric charge density is assumed uniform, the integral (\ref{eq:7})
is proportional to this apparent volume, the volume of the effective integration region.

\section{The actual volume of a spherical particle in motion}

Consider a very small spherical particle of radius $a'$ and uniform electric charge density $\rho'$, at rest in
the reference frame $K'$. The particle is at a distance $d$ from the origin, on the $Ox'$ axis.
The volume of the sphere is $4 \pi a'^3 / 3$, and the total electric charge of the particle is $Q = \rho'  4 \pi a'^3 / 3$.
The center of the sphere, with coordinates 
\begin{equation}
(x'_\alpha, y'_\alpha, z'_\alpha) = (d, 0, 0), 
\label{eq:alpha}
\end{equation}
belongs to worldline $\alpha$.
Inside this particle we consider a second point, also at rest in the reference frame $K'$, with coordinates 
\begin{equation}
(x'_\beta, y'_\beta, z'_\beta) = (d + \xi, \eta, \zeta), 
\label{eq:beta}
\end{equation}
that belongs to worldline $\beta$.
The region with a non-zero electric charge density is given by
\begin{equation}
( x' - d )^2 + y'^2 + z'^2 \leq a'^2,
\label{eq:sphere2}
\end{equation}
which, in Minkowski space, is a worldtube parallel to the $O i c t'$ axis.

Consider now that the particle is in motion along the $Ox$ axis of another reference frame $K$, 
with a uniform velocity $\overrightarrow{v} = (v, 0, 0)$. 
The particle's proper reference frame $K'$ moves with the same velocity relative to this stationary reference frame $K$. 
The origins of the two reference frames coincide when $t = t'= 0$.
We use the Lorentz transformation
\begin{equation}
x'= \frac{x - v t}{\sqrt{1 - v^2 / c^2}}, \ \ \ y'= y, \ \ \ z'= z, \ \ \ t'= \frac{t - v x / c^2}{\sqrt{1 - v^2 / c^2}},
\end{equation}
to make substitutions into (\ref{eq:alpha}), (\ref{eq:beta}), and (\ref{eq:sphere2}). 

In the stationary reference frame $K$ the $\alpha$ worldline is given by
\begin{equation}
 (x_\alpha, y_\alpha, z_\alpha) = (d \sqrt{1 - v^2 / c^2} +  v t, 0, 0),
\label{eq:alpha2}
\end{equation}
and the $\beta$ worldline is given by
\begin{equation}
(x_\beta, y_\beta, z_\beta) = ((d + \xi) \sqrt{1 - v^2 / c^2} + v t, \eta, \zeta).
\label{eq:beta2}
\end{equation}
The region with a non-zero electric charge density is given by
\begin{equation}
\frac{( x - v t - d \sqrt{1 - v^2 / c^2})^2}{1 - v^2 / c^2} + y^2 + z^2 \leq a'^2.
\label{eq:worldtube}
\end{equation}

Through the field point $P$ at $(x_P, y_P, z_P, i c t_P)$ we draw the hyperplane of constant time $t = t_P$. The intersection of this
hyperplane with the worldtube of the particle (\ref{eq:worldtube}) produces a space region described by
\begin{equation}
\frac{( x - v t_P - d \sqrt{1 - v^2 / c^2})^2}{1 - v^2 / c^2} + y^2 + z^2 \leq a'^2.
\label{Lorentzellipsoid}
\end{equation}
This is the actual (instantaneous) integration volume, a Lorentz ellipsoid with axes $a' \sqrt{1 - v^2 / c^2}$, $a'$, and $a'$. 
The center of the ellipsoid, the point with coordinates $(v t_P + d \sqrt{1 - v^2 / c^2}, 0, 0)$, 
belongs to worldline $\alpha$.
The volume of the ellipsoid is $4 \pi a'^3 \sqrt{1 - v^2 / c^2} / 3$, 
and the total electric charge of the particle is $Q = \rho  4 \pi a'^3 \sqrt{1 - v^2 / c^2} / 3$.

The invariance of electric charge under a Lorentz boost, expressed by $\rho' = \rho  \sqrt{1 - v^2 / c^2}$, 
is consistent with the fact that \lq\lq total charge could be measured by a counting operation
which is presumably also an invariant\rq\rq. \cite{Panofsky} 
Unlike the invariance of electric charge, the Doppler factor in the LW potentials
\lq\lq has nothing whatever to do with special relativity or Lorentz contraction\rq\rq. \cite{Griffiths4ed} 

\section{The apparent volume of a spherical particle in motion}

Through the field point $P$ at $(x_P, y_P, z_P, i c t_P)$ we draw the retarded lightcone, a hypersurface described by the equation
\begin{equation}
( x - x_P )^2 + ( y - y_P )^2 + ( z - z_P )^2 - c^2 ( t - t_P )^2 = 0,
\label{eq:lightcone}
\end{equation}
which, since $t < t_P$ for a retarded time, we can also write as
\begin{equation}
c ( t_P - t ) = \sqrt{( x - x_P )^2 + ( y - y_P )^2 + ( z - z_P )^2}.
\label{eq:lightcone2}
\end{equation}
The intersection of the retarded
lightcone (\ref{eq:lightcone2}) with the worldtube of the particle (\ref{eq:worldtube}) 
is a spacetime region whose projection on the 
hyperplane of constant time $t = t_P$ gives the apparent integration volume.

In order to find the ratio of the apparent volume to the actual volume, 
we compare the spatial separation of two points from the apparent volume 
with the spatial separation of two corresponding points from the actual volume. 

Worldline $\alpha$ intersects the hyperplane of constant time $t = t_P$ at $A$
\begin{equation}
 (x_A, y_A, z_A, i c t_A) = (d \sqrt{1 - v^2 / c^2} +  v t_P, 0, 0, i c t_P),
\label{eq:A}
\end{equation}
and worldline $\beta$ intersects the same hyperplane at $B$
\begin{equation}
(x_B, y_B, z_B, i c t_B) = ((d + \xi) \sqrt{1 - v^2 / c^2} + v t_P, \eta, \zeta, i c t_P).
\label{eq:B}
\end{equation}
Worldline $\alpha$ intersects the the retarded lightcone (\ref{eq:lightcone2}) at $C$
\begin{equation}
 (x_C, y_C, z_C, i c t_C) = (d \sqrt{1 - v^2 / c^2} +  v t_C, 0, 0, i c t_C),
\label{eq:C}
\end{equation}
and worldline $\beta$ intersects the same lightcone at $D$
\begin{equation}
(x_D, y_D, z_D, i c t_D) = ((d + \xi) \sqrt{1 - v^2 / c^2} + v t_D, \eta, \zeta, i c t_D).
\label{eq:D}
\end{equation}
The retarded point $C$ is projected on the hyperplane of constant time $t = t_P$ at $E$
\begin{equation}
 (x_E, y_E, z_E, i c t_E) = (d \sqrt{1 - v^2 / c^2} +  v t_C, 0, 0, i c t_P),
\label{eq:E}
\end{equation}
and the retarded point $D$ is projected on the same hyperplane at $F$
\begin{equation}
(x_F, y_F, z_F, i c t_F) = ((d + \xi) \sqrt{1 - v^2 / c^2} + v t_D, \eta, \zeta, i c t_P).
\label{eq:F}
\end{equation}

\begin{figure}[h!]
\begin{center}
\includegraphics[height=14.535cm]{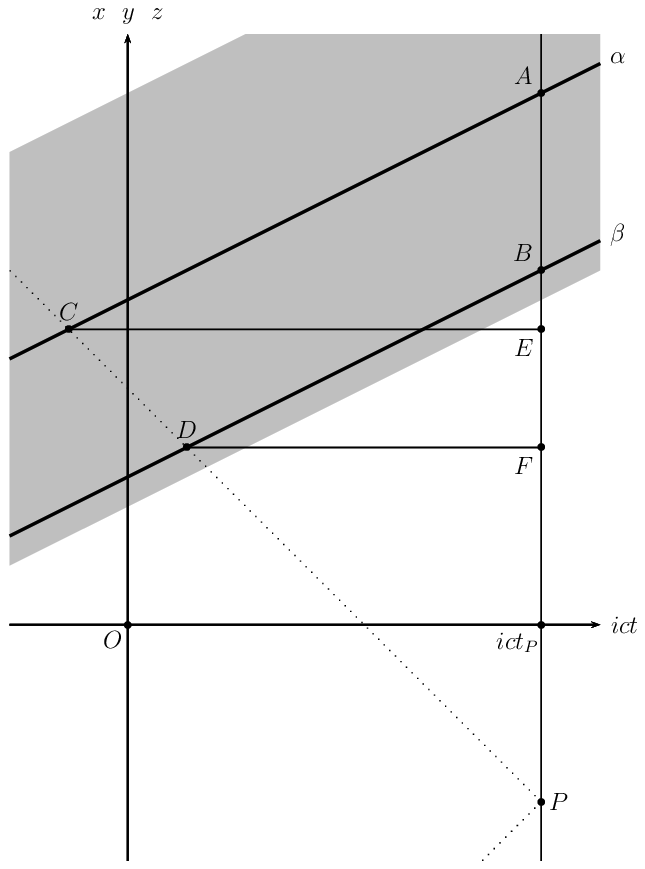}
\caption{An electrically charged extended particle in motion (gray area),
and the field point $P$.}
\label{fig:1}
\end{center}
\end{figure}

As shown in Figure 1, the spatial separation of points $A$ and $B$, inside the actual volume, is
\begin{equation}
(x_B, y_B, z_B) -  (x_A, y_A, z_A) = (\xi \sqrt{1 - v^2 / c^2}, \eta, \zeta),
\label{eq:PB}
\end{equation}
while the spatial separation of the corresponding points $E$ and $F$, inside the apparent volume, is
\begin{equation}
(x_F, y_F, z_F) -  (x_E, y_E, z_E) = (\xi \sqrt{1 - v^2 / c^2} + v (t_D - t_C), \eta, \zeta).
\label{eq:EF}
\end{equation}
The ratio $(x_F - x_E)/(x_B - x_A)$, equal to the ratio of the apparent to the actual volume, 
is responsible for the apparition of the Doppler factor in the LW potentials. 

\subsection{Electric charge density in radial motion}

From a pedagogical point of view, it is helpful to start by analyzing the simpler case of an extended 
particle in radial motion. \cite{Griffiths4ed,FeynmanII,Dikshit}
In this situation the fieldpoint is on the $Ox$ axis, and therefore $y_P = z_P = 0$.
The center of the retarded electric charge density is at point $C$, and 
$\overrightarrow{R} = \overrightarrow{r_P} - \overrightarrow{r_C} = ( x_P - x_C , 0 , 0 )$. 
When $x_C > x_P$, the unit vector $\hat{R}$ points in the negative $x$ direction, $\hat{R} = (-1, 0, 0)$,
and then $1 - \hat{R} \cdot \overrightarrow{v} / c = 1 + v / c$.
When $x_C < x_P$, the unit vector $\hat{R}$ points in the positive $x$ direction, $\hat{R} = (1, 0, 0)$,
and then $1 - \hat{R} \cdot \overrightarrow{v} / c = 1 - v / c$.

From (\ref{eq:C}) we know that $y_C = z_C = 0$, and equation (\ref{eq:lightcone2}) becomes
\begin{equation}
c ( t_P - t_C ) = \sqrt{( x_C - x_P )^2},
\label{eq:lightcone3}
\end{equation}
where 
\begin{equation}
 x_C = d \sqrt{1 - v^2 / c^2} +  v t_C.
\label{eq:xC}
\end{equation}

When $x_C > x_P$, from (\ref{eq:lightcone3}) we get
\begin{equation}
c ( t_P - t_C ) = x_C - x_P,
\end{equation}
which, together with (\ref{eq:xC}), allows us to find
\begin{eqnarray}
t_C = \frac{x_P + c t_P - d \sqrt{1 - v^2 / c^2}}{c + v}, \\
x_C = \frac{d \sqrt{1 - v^2 / c^2} +  v t_P + x_P v / c}{1 + v / c} = \frac{x_A + x_P v / c}{1 + v / c}.
\label{eq:xCgreater}
\end{eqnarray}

When $x_C < x_P$, from (\ref{eq:lightcone3}) we get
\begin{equation}
c ( t_P - t_C ) = x_P - x_C,
\end{equation}
which, together with (\ref{eq:xC}), allows us to find
\begin{eqnarray}
t_C = \frac{- x_P + c t_P + d \sqrt{1 - v^2 / c^2}}{c - v}, \\
x_C = \frac{d \sqrt{1 - v^2 / c^2} +  v t_P - x_P v / c}{1 - v / c} = \frac{x_A - x_P v / c}{1 - v / c}.
\label{eq:xCless}
\end{eqnarray}

From (\ref{eq:D}) we know that $y_D = \eta$, $z_D = \zeta$, and equation (\ref{eq:lightcone2}) becomes
\begin{equation}
c ( t_P - t_D ) = \sqrt{( x_D - x_P )^2 + \eta^2 + \zeta^2} 
\approx \sqrt{( x_D - x_P )^2} \left( 1 + \frac{\eta^2 + \zeta^2}{2 ( x_D - x_P )^2} \right),
\label{eq:lightcone4}
\end{equation}
where 
\begin{equation}
 x_D = (d + \xi) \sqrt{1 - v^2 / c^2} +  v t_D.
\label{eq:xD}
\end{equation}
The last part of equation (\ref{eq:lightcone4}) comes from a Taylor series expansion 
of the square root function, $\sqrt{1 + \epsilon} \approx 1 + \epsilon / 2$ for $\epsilon << 1$.
Since the extended particle is of very small size, in order to calculate the spatial separations (\ref{eq:PB}) and (\ref{eq:EF}) 
we only need the first order contributions in $\xi$, $\eta$, and $\zeta$.
In this approximation (\ref{eq:lightcone4}) becomes 
\begin{equation}
c ( t_P - t_D ) = \sqrt{( x_D - x_P )^2}.
\label{eq:lightcone5}
\end{equation}

When $x_D > x_P$ (implying $x_C > x_P$, since points $C$ and $D$ are very close to each other), from (\ref{eq:lightcone5}) we get
\begin{equation}
c ( t_P - t_D ) = x_D - x_P,
\end{equation}
which, together with (\ref{eq:xD}), allows us to find
\begin{eqnarray}
t_D = \frac{x_P + c t_P - (d + \xi) \sqrt{1 - v^2 / c^2}}{c + v}, \\
x_D = \frac{(d + \xi) \sqrt{1 - v^2 / c^2} +  v t_P + x_P v / c}{1 + v / c} = \frac{x_B + x_P v / c}{1 + v / c}.
\label{eq:xDgreater}
\end{eqnarray}

When $x_D < x_P$  (implying $x_C < x_P$), from (\ref{eq:lightcone5}) we get
\begin{equation}
c ( t_P - t_D ) = x_P - x_D,
\end{equation}
which, together with (\ref{eq:xD}), allows us to find
\begin{eqnarray}
t_D = \frac{- x_P + c t_P + (d + \xi) \sqrt{1 - v^2 / c^2}}{c - v}, \\
x_D = \frac{(d + \xi) \sqrt{1 - v^2 / c^2} +  v t_P - x_P v / c}{1 - v / c} = \frac{x_B - x_P v / c}{1 - v / c}.
\label{eq:xDless}
\end{eqnarray}

Since $x_E = x_C$ and $x_F = x_D$, it is now clear that in both cases ( $x_C > x_P$ or $x_C < x_P$ ) we arrive at the Doppler factor
\begin{equation}
\frac{\textrm{apparent volume}}{\textrm{actual volume}} 
= \frac{x_F - x_E}{x_B - x_A} 
= \frac{x_D - x_C}{x_B - x_A} 
= \frac{1}{1 - \hat{R} \cdot \overrightarrow{v} / c}.
\label{eq:Doppler0}
\end{equation}

\subsection{Electric charge density in non-radial motion}

In this general situation the fieldpoint $P$ can be anywhere.
The retarded electric charge density is non-zero in a very small neighborhood of point $C$, and 
$\overrightarrow{R} = \overrightarrow{r_P} - \overrightarrow{r_C} = ( x_P - x_C , y_P , z_P )$. 
The radial unit vector is 
\begin{equation}
\hat{R} = \frac{( x_P - x_C , y_P , z_P )}{\sqrt{(x_P - x_C)^2 + y_P^2 + z_P^2}},
\label{eq:Rhat}
\end{equation}
and the denominator of the Doppler factor is
\begin{equation}
1 - \frac{\hat{R} \cdot \overrightarrow{v}}{c} = 1 - \frac{v ( x_P - x_C )}{c \sqrt{(x_P - x_C)^2 + y_P^2 + z_P^2}}.
\label{eq:Doppler}
\end{equation}

For point $C$ equation (\ref{eq:lightcone2}) becomes
\begin{equation}
c ( t_P - t_C ) = \sqrt{( x_C - x_P )^2 + y_P^2 + z_P^2},
\label{eq:lightcone6}
\end{equation}
and if we substitute (\ref{eq:xC}) into (\ref{eq:lightcone6}) we obtain a quadratic equation in $t_C$. 
One solution is the retarded time that we keep, the other is the advanced time that we discard.
However, unlike in the simpler case of radial motion, in the general case the discriminant of the quadratic equation
is no longer a perfect square, and the solution does not look very nice.

For point $D$ equation (\ref{eq:lightcone2}) becomes
\begin{equation}
c ( t_P - t_D ) = \sqrt{( x_D - x_P )^2 + ( \eta - y_P )^2 + ( \zeta - z_P )^2},
\label{eq:lightcone7}
\end{equation}
and if we substitute (\ref{eq:xD}) into (\ref{eq:lightcone7}) we obtain a quadratic equation in $t_D$.
We notice that we don't have to actually find the exact solution of this quadratic equation, since we are only interested in the
first order approximation in $\xi$, $\eta$, $\zeta$. For this reason we decide to consider $\xi$, $\eta$, $\zeta$ as free parameters,
and we look at the exact solutions $t_D(\xi, \eta, \zeta)$ and $x_D(\xi, \eta, \zeta)$ as functions of these parameters. Since
$x_D(0, 0, 0) = x_C$ and $t_D(0, 0, 0) = t_C$, we can expand $x_D$ in a Taylor series around $x_C$, 
keeping only the first order contributions in $\xi$, $\eta$, $\zeta$, and we obtain
\begin{equation}
x_D(\xi, \eta, \zeta) = x_C + \left. \frac{\partial x_D}{\partial \xi}\right\vert_C \xi
+ \left. \frac{\partial x_D}{\partial \eta}\right\vert_C \eta + \left. \frac{\partial x_D}{\partial \zeta}\right\vert_C \zeta.
\label{eq:taylor}
\end{equation}

By applying partial derivatives to (\ref{eq:xD}) we get
\begin{eqnarray}
\frac{\partial x_D}{\partial \xi} = \sqrt{1 - \frac{v^2}{c^2}} + v \frac{\partial t_D}{\partial \xi}, \label{eq:xi1} \\
\frac{\partial x_D}{\partial \eta} = v \frac{\partial t_D}{\partial \eta}, \label{eq:eta1} \\
\frac{\partial x_D}{\partial \zeta} = v \frac{\partial t_D}{\partial \zeta}, \label{eq:zeta1}
\end{eqnarray}
and by applying partial derivatives to (\ref{eq:lightcone7}) we get
\begin{eqnarray}
- c \frac{\partial t_D}{\partial \xi} 
= \frac{( x_D - x_P ) \frac{\partial x_D}{\partial \xi}}{\sqrt{( x_D - x_P )^2 + ( \eta - y_P )^2 + ( \zeta - z_P )^2}}, \label{eq:xi2} \\
- c \frac{\partial t_D}{\partial \eta} 
= \frac{( x_D - x_P ) \frac{\partial x_D}{\partial \eta} + \eta - y_P}{\sqrt{( x_D - x_P )^2 + ( \eta - y_P )^2 + ( \zeta - z_P )^2}}, \label{eq:eta2} \\
- c \frac{\partial t_D}{\partial \zeta}
= \frac{( x_D - x_P ) \frac{\partial x_D}{\partial \zeta} + \zeta - z_P}{\sqrt{( x_D - x_P )^2 + ( \eta - y_P )^2 + ( \zeta - z_P )^2}}. \label{eq:zeta2}
\end{eqnarray}

From (\ref{eq:xi1}) and (\ref{eq:xi2}) we find that
\begin{equation}
\frac{\partial x_D}{\partial \xi} 
= \frac{\frac{c}{v} \sqrt{1 - \frac{v^2}{c^2}} \sqrt{( x_D - x_P )^2 + ( \eta - y_P )^2 + ( \zeta - z_P )^2}}
{x_D - x_P + \frac{c}{v} \sqrt{( x_D - x_P )^2 + ( \eta - y_P )^2 + ( \zeta - z_P )^2}}, 
\end{equation}
which gives
\begin{equation}
\left. \frac{\partial x_D}{\partial \xi}\right\vert_C 
= \frac{\frac{c}{v} \sqrt{1 - \frac{v^2}{c^2}} \sqrt{( x_C - x_P )^2 + y_P^2 + z_P^2}}
{x_C - x_P + \frac{c}{v} \sqrt{( x_C - x_P )^2 + y_P^2 + z_P^2}}.
\end{equation}

From (\ref{eq:eta1}) and (\ref{eq:eta2}) we find that
\begin{equation}
\frac{\partial x_D}{\partial \eta} 
= \frac{y_P - \eta}{x_D - x_P + \frac{c}{v} \sqrt{( x_D - x_P )^2 + ( \eta - y_P )^2 + ( \zeta - z_P )^2}}, 
\end{equation}
which gives
\begin{equation}
\left. \frac{\partial x_D}{\partial \eta}\right\vert_C 
= \frac{y_P}{x_C - x_P + \frac{c}{v} \sqrt{( x_C - x_P )^2 + y_P^2 + z_P^2}}.
\end{equation}

From (\ref{eq:zeta1}) and (\ref{eq:zeta2}) we find that
\begin{equation}
\frac{\partial x_D}{\partial \zeta} 
= \frac{z_P - \zeta}{x_D - x_P + \frac{c}{v} \sqrt{( x_D - x_P )^2 + ( \eta - y_P )^2 + ( \zeta - z_P )^2}}, 
\end{equation}
which gives
\begin{equation}
\left. \frac{\partial x_D}{\partial \zeta}\right\vert_C 
= \frac{z_P}{x_C - x_P + \frac{c}{v} \sqrt{( x_C - x_P )^2 + y_P^2 + z_P^2}}.
\end{equation}

With these partial derivatives, equation (\ref{eq:taylor}) becomes
\begin{equation}
x_D = x_C + \frac{\frac{c}{v} \sqrt{1 - \frac{v^2}{c^2}} \sqrt{( x_C - x_P )^2 + y_P^2 + z_P^2} \xi + y_P \eta + z_P \zeta}
{x_C - x_P + \frac{c}{v} \sqrt{( x_C - x_P )^2 + y_P^2 + z_P^2}},
\label{eq:taylor2}
\end{equation}
and since $x_B - x_A = \xi \sqrt{1 - v^2 / c^2}$, we have
\begin{equation}
\frac{x_D - x_C}{x_B - x_A} 
= \frac{\frac{c}{v} \sqrt{1 - \frac{v^2}{c^2}} \sqrt{( x_C - x_P )^2 + y_P^2 + z_P^2} \xi + y_P \eta + z_P \zeta}
{\xi \sqrt{1 - \frac{v^2}{c^2}} \left( x_C - x_P + \frac{c}{v} \sqrt{( x_C - x_P )^2 + y_P^2 + z_P^2} \right) }.
\label{eq:Doppler2}
\end{equation}

We now make use of the infinitesimal nature of $\eta$ and $\zeta$, and we evaluate (\ref{eq:Doppler2}) 
in the limit $\eta \to 0$ and $\zeta \to 0$. Surprisingly, $\xi$ cancels out and we don't have to worry about its limit.
From this perspective, the Doppler factor in the LW potentials is seen as \lq\lq a zeroth-order contribution\rq\rq. \cite{Aguirregabiria} We have
\begin{equation}
\lim \limits_{\substack{\eta \to 0 \\ \zeta \to 0}} \frac{x_D - x_C}{x_B - x_A}
= \frac{\frac{c}{v} \sqrt{( x_C - x_P )^2 + y_P^2 + z_P^2}}
{x_C - x_P + \frac{c}{v} \sqrt{( x_C - x_P )^2 + y_P^2 + z_P^2}}.
\label{eq:Doppler3}
\end{equation}

Based on (\ref{eq:Doppler}) and (\ref{eq:Doppler3}), we finally arrive at the Doppler factor
\begin{equation}
\frac{\textrm{apparent volume}}{\textrm{actual volume}} 
= \frac{1}{1 - \frac{v}{c} \frac{x_P - x_C}{\sqrt{( x_C - x_P )^2 + y_P^2 + z_P^2}}}
= \frac{1}{1 - \hat{R} \cdot \overrightarrow{v} / c}.
\label{eq:Doppler4}
\end{equation}

\section{The transition to an electrically charged point particle}

It is generally assumed that the derivation of the Doppler factor in the LW potentials, 
given in the case of a continuous electric charge distribution,
also applies in the case of an electrically charged point particle.
This assumption seems natural, since the LW potentials of a point charge 
are exactly the same as those of a continuous charge distribution localized in a vanishingly small 3D volume,
as long as the point particle and the extended particle have the same electric charge and velocity.
The widespread nature of the assumption is demonstrated by its apparition in some important physics textbooks.
For example, Griffiths writes:
\lq\lq Because this correction factor makes no reference to the {\it size} of the particle, it is every bit as significant for a point charge as for an extended charge.\rq\rq \cite{Griffiths4ed} 
and Feynman writes:
\lq\lq Finally, since the \lq size\rq \ of the charge $q$ doesn’t enter into the final result, the same result holds when we let the charge shrink to any size - even to a point.\rq\rq \cite{FeynmanII} 

We respectfully disagree with these statements. We believe that the assumption is not justified, and this is the reason: 
While in the case on an extended particle we can have two worldlines $\alpha$ and $\beta$ inside the particle, 
and the derivation of the Doppler factor in the LW potentials
proceeds as shown, in the case of a point particle we only have one worldline. The previous reasoning no longer applies,
the ratios (\ref{eq:Doppler0}) and (\ref{eq:Doppler4}) no longer exist. 
Other authors have also voiced concerns. 
Joel Franklin calls the assumption \lq\lq a hard sell\rq\rq \cite{JoelFranklin}, and Vesselin Petkov writes:
\lq\lq If this were the case, the explanation of the physical origin of the Li\'{e}nard-Wiechert potentials would not make sense.\rq\rq \cite{Petkov}
The above mentioned assumption has survived for so long only because the conceptual difference between a diameter of infinitesimal length
and a point of zero size is quite elusive. 

\section{Electrically charged point particle in motion}

We have seen that, for an extended particle, the Doppler factor in the LW potentials has a geometrical origin,
this factor being the result of a geometrical derivation performed in Minkowski space.
We cannot use the same proof, based on the two worldlines $\alpha$ and $\beta$, for a point particle.
Nonetheless, the question is, can we still find the geometrical origin of the Doppler factor in this later case?

In this quest, we start by noticing that \cite{Landau4ed}
\begin{equation}
\frac{\partial t_{ret}}{\partial t} = \frac{1}{1 - \hat{R} \cdot \overrightarrow{v} / c},
\label{eq:65}
\end{equation}
where $t$ is the time at the field point ($t_P$ in our notation), 
and $t_{ret}$ is the retarded time at the source charge ($t_C$ in our notation).
Equation (\ref{eq:4}) becomes
\begin{equation}
\mathbf{\Phi}_P = \frac{Q}{R} \left( \frac{v_x}{c}, \frac{v_y}{c}, \frac{v_z}{c}, i \right) \frac{\partial t_{ret}}{\partial t}.
\label{eq:4new}
\end{equation}

In order to understand the geometrical origin of the Doppler factor (\ref{eq:65}) we need a geometrical representation of (\ref{eq:4new}).
Consider an electrically charged point particle in motion, with a retarded velocity $\overrightarrow{v} = (v_x, v_y, v_z)$, 
and a field point $P$ with coordinates $(x_P, y_P, z_P, i c t_P)$.
As shown in Figure 2, through the field point $P$  
we draw the retarded lightcone that intersects the 
worldline of the source particle at point $C$ with coordinates $(x_C, y_C, z_C, i c t_C)$. 
Point $C$ is projected on the time axis at $E$.
In order to calculate the partial derivative of $t_{ret}$ with respect to $t$ we keep the $x_P$, $y_P$, $z_P$ 
coordinates of the field point $P$ fixed, and we increase the time $t_P$ by an infinitesimal amount $\delta t$,
obtaining in this way the new field point $B$ with coordinates $(x_B, y_B, z_B, i c t_B) = (x_P, y_P, z_P, i c t_P + i c \delta t)$.
Through the field point $B$ we draw the retarded lightcone that intersects the 
worldline of the source particle at point $D$ with coordinates $(x_D, y_D, z_D, i c t_D)$. 
Point $D$ is projected on the time axis at $F$.
The variation of the retarded time is
$\delta t_{ret} = t_D - t_C$, and the Doppler factor (\ref{eq:65}) is
\begin{equation}
\frac{\partial t_{ret}}{\partial t} = \frac{\delta t_{ret}}{\delta t} = \frac{t_D - t_C}{t_B - t_P}.
\label{eq:67}
\end{equation}
It is as if the point particle, devoid of a spatial volume, has instead gained a temporal extension,
and now the ratio of these temporal dimensions is responsible for the apparition of the Doppler factor.

\begin{figure}[h!]
\begin{center}
\includegraphics[height=9.525cm]{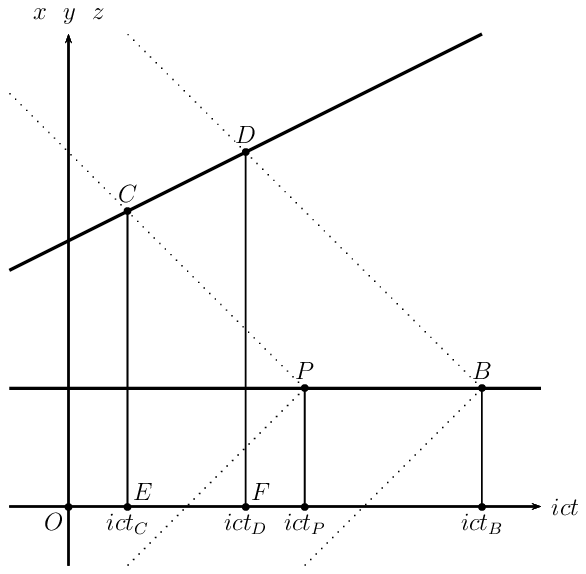}
\caption{An electrically charged point particle in motion (worldline CD),
and the field points $P$ and $B$.}
\label{fig:2}
\end{center}
\end{figure}

We notice that the right side of (\ref{eq:67}) is the temporal component of a Minkowski four-vector
\begin{multline}
\frac{\overrightarrow{CD}}{PB} = \frac{(x_D - x_C, y_D - y_C, z_D - z_C, i c t_D - i c t_C)}{i c t_B - i c t_P} \\
= \frac{(v_x \delta t_{ret}, v_y \delta t_{ret}, v_z \delta t_{ret}, i c \delta t_{ret})}{i c \delta t}
= \frac{1}{i} \left( \frac{v_x}{c}, \frac{v_y}{c}, \frac{v_z}{c}, i \right) \frac{\partial t_{ret}}{\partial t}.
\end{multline}
This allows us to write (\ref{eq:4new}) as
\begin{equation}
\mathbf{\Phi}_P = \frac{i Q}{R} \frac{\overrightarrow{CD}}{PB}.
\label{eq:69}
\end{equation}
At the same time, $\overrightarrow{CP} = (\overrightarrow{R}, i R)$ and  $\overrightarrow{PB} = (\overrightarrow{0}, i c \delta t)$,
where the length of $\overrightarrow{PB}$ is $PB = i c \delta t$. Since $\overrightarrow{CP} \cdot \overrightarrow{PB} = i R\ PB$,
we can write (\ref{eq:69}) as
\begin{equation}
\mathbf{\Phi}_P = - \frac{Q\  \overrightarrow{CD}}{\overrightarrow{CP} \cdot \overrightarrow{PB}}.
\label{eq:69new}
\end{equation}

We recognize that the displacement four-vector $\overrightarrow{CD}$ from (\ref{eq:69new}) is closely related to the 
four-velocity $\mathbf{V}_C$ from (\ref{eq:3}), because $\mathbf{V}_C = \overrightarrow{CD} / \delta\tau_{ret}$, where 
$\delta\tau_{ret}$ is an infinitesimal retarded proper time interval 
and the length of $\overrightarrow{CD}$ is $CD = i c \delta\tau_{ret}$.
Since $\mathbf{X}_P - \mathbf{X}_C = \overrightarrow{CP}$, we can write (\ref{eq:3}) as
\begin{equation}
\mathbf{\Phi}_P = - \frac{Q\  \overrightarrow{CD}}{\overrightarrow{CP} \cdot \overrightarrow{CD}}.
\label{eq:3new}
\end{equation}

Comparing (\ref{eq:3new}) with (\ref{eq:69new}), a very important equation emerges
\begin{equation}
\overrightarrow{CP} \cdot \overrightarrow{PB} = \overrightarrow{CP} \cdot \overrightarrow{CD}.
\label{eq:fokker}
\end{equation}
This equation is true whenever two infinitesimal segments (in our case $PB$ and $CD$)
have their endpoints connected by light signals.
The same equation (\ref{eq:fokker}) was used by Fokker \cite{fokker}
who, in his variational method, assumed that the 
electromagnetic interaction takes place between such
corresponding effective elements (\lq\lq entsprechended effektiven Elementen\rq\rq),
segments of infinitesimal length on the worldlines of the electrically charged particles,
with null spacetime intervals between their corresponding endpoints. 
In Fokker's notation, the relation (\ref{eq:fokker}) is written as $(R \cdot dx) = (R \cdot dy)$.

Next assume that at the field point $P$ we have an electrically charged particle at rest, with electric charge $q$. 
The infinitesimal segment $PB$ lies on the worldline of this test charge. The electromagnetic interaction is described by the
term $(q/c) \sqrt{1 - v_P^2/c^2}\,\mathbf{\Phi}_P \cdot \mathbf{V}_P$ in the relativistic Lagrangian \cite{Jackson}, 
and by the term $(q/c)\,\mathbf{\Phi}_P \cdot \mathbf{V}_P\,d\tau  = (q/c)\,\mathbf{\Phi}_P \cdot \mathbf{dX}_P$ 
in the integral of the action,
where $d\tau = \sqrt{1 - v_P^2/c^2} dt$ is an infinitesimal proper time interval
and $\mathbf{dX}_P$ is the corresponding infinitesimal variation of the position four-vector $\mathbf{X}_P$.
Jackson has an additional minus sign in his expression, because he is using 
a Minkowski metric tensor of signature $( - , - , - , + )$,
while we are using Minkowski's imaginary time formalism, equivalent to 
a Minkowski metric tensor of signature $( + , + , + , - )$.
Since in our case $\mathbf{dX}_P = \overrightarrow{PB}$, we end up with
\begin{equation}
- \frac{Q\  q}{c} \frac{\overrightarrow{CD} \cdot \overrightarrow{PB}}{\overrightarrow{CP} \cdot \overrightarrow{CD}},
\label{eq:action}
\end{equation}
in the relativistic action.
The expression (\ref{eq:action}) is a Lorentz invariant, and it describes the electromagnetic
interaction even when the test charge $PB$ is not at rest.
Due to symmetry considerations (related to the action and reaction principle), 
Fokker has also added to (\ref{eq:action}) the contribution of the advanced potential.

One way to reveal the geometrical origin of the Doppler factor in the LW potentials 
is to say that electrically charged point particles are points only in 3D space,
while in Minkowski space they are not points, but infinitesimal length elements along the worldlines of the particles.
The electromagnetic interaction, as shown by the expression of the relativistic action, 
happens between infinitesimal worldline segments that have their end points connected by light signals.

This interaction model, first proposed by Fokker, was also independently discovered by Galeriu \cite{galeriuAAAD},
based on the following argument:
Consider a source charge at rest and a test charge in uniform motion.
In the expression of the electromagnetic four-force 
we have an explicit dependence on the velocity of the test particle, 
and an implicit dependence on the velocity of the source particle (through the chosen reference frame). 
However, \lq\lq from a {\it geometrical} point of view, a point in Minkowski space
is just a fixed point - it does not have a velocity!\rq\rq \cite{galeriuAAAD} 
The four-force acting on a point particle must be replaced by a linear four-force density 
acting on an infinitesimal segment on the particle's worldline. \cite{galeriuArXiv}
This paradigm shift is mathematically possible because, as noticed by Costa de Beauregard \cite{costa1},
we have a perfect isomorphism between the equation of motion for a relativistic point particle
and the static equilibrium condition for an elastic string.

It is very likely that Minkowski himself was on the verge of discovering the possibility of replacing 
the interacting material point particles with corresponding infinitesimal segments on 
the particles' worldlines, given the words \lq\lq Let BC be an infinitely small element of the worldline of F; 
further let B* be the light point of B, C* be the light point of C on the worldline of F* [...]\rq\rq that he used 
when describing his first theory of gravitational interaction \cite{minkowski2}. These words 
would have easily escaped our attention, were it not for
a matching diagram drawn by Scott Walter. \cite{ScottWalter}   

Another way to reveal the geometrical origin of the Doppler factor in the LW potentials
is to talk about the \lq\lq thickness\rq\rq of the lightcone. There are two possible ways of drawing the thick lightcones.
One could hold the $PB$ segment of the test charge constant, draw the lightcones with
vertices at $P$ and $B$, and notice that the velocity of the source charge will change the length of the 
intersection segment $CD$ that the source charge has inside the thick lightcone.
Or one could hold the $CD$ segment of the source charge constant, draw the lightcones with
vertices at $C$ and $D$, and notice that the velocity of the test charge will change the length of the
intersection segment $PB$ that the test charge has inside the thick lightcone. 

The second method was used by Nicholas Wheeler and Kevin Brown.
After deriving the LW potentials, Nicholas Wheeler draws a thick lightcone and writes:
\lq\lq If the lightcone had \lq thickness\rq \ then the presence
of the Doppler factor [...] could be understood qualitatively to
result from the relatively \lq longer look\rq \ that the field point gets at
approaching charges, the relatively \lq briefer look\rq \ at receding charges.\rq\rq
\cite{NicholasWheeler} 
Analyzing the LW potentials in the simpler radial case, 
Kevin Brown draws a thick lightcone and writes: \lq\lq The light cone is shown with a non-zero thickness 
to illustrate that the duration of time spent by each particle as it passes through the light cone depends on the speed of the particle. [...] In general, the duration 
of coordinate time spent by a point-like particle in the light cone shell is proportional to $1/(1 + v/c)$.\rq\rq \cite{brown} 

\section{The relativistic Doppler effect factor}

The algebraic structure of the Doppler factor (\ref{eq:65}) closely mirrors 
the algebraic structure of the classical Doppler effect factor \cite{Neipp}, 
and this was most likely the reason for the name given to this factor in the LW potentials. 
However, historically, the exact relationship between the two factors was not very well understood.
For example, O'Rahilly mentions that \lq\lq Many writers regard the factor $1/(1 - v_R'/c)$ 
as somehow connected with Doppler's principle. 
Heaviside [...] says the Li\'{e}nard potential is \lq dopplerised.\rq \ \rq\rq
\cite{ORahilly}
Now, based on our geometrical interpretation,
we can understand the relationship between the Doppler factor in the LW potentials and the classical Doppler effect factor. 
The missing link is the relativistic Doppler effect factor.

Indeed, we can use the Minkowski diagram shown in Figure 2 in order to discuss the relativistic Doppler effect \cite{Synge1963,TevianDray}.
Worldline $CD$ represents the source of the electromagnetic wave, and worldline $PB$ represents the detector. Suppose that
flashes of light are emitted once per period. The flashes emitted at $C$ and $D$ are received at $P$ and $B$. 
The spacetime interval $CD = i c T'$ gives the period $T'$ of the wave in the proper reference frame of the source,
while the spacetime interval $PB = i c T$ gives the period $T$ of the wave in the reference frame of the detector.
The relativistic Doppler effect factor is 
the ratio of the frequency $f$ measured by the detector
to the frequency $f'$ produced by the source:
\begin{equation}
\frac{f}{f'} = \frac{T'}{T} = \frac{CD}{PB} = \frac{CD}{EF} \frac{EF}{PB} = \sqrt{1 - \frac{v^2}{c^2}} \frac{t_D - t_C}{t_B - t_P}.
\label{eq:relativisticDoppler}
\end{equation}
The only difference between the relativistic Doppler effect factor (\ref{eq:relativisticDoppler})
and the Doppler factor in the LW potentials (\ref{eq:67}) is the
$\sqrt{1 - v^2 / c^2}$ factor. This additional factor, responsible for the transverse Doppler effect,
represents the relativistic time dilation. \cite{ataman}

\section{Concluding remarks}

For an electrically charged extended particle, we have derived the Doppler factor in the LW potentials
with the help of intersections and projections, using analytical geometry methods. 
In Minkowski space nothing moves, 
nothing changes shape, and as a result our geometrical derivation is very intuitive.
For an electrically charged point particle, we have discussed how the Doppler factor in the LW potentials
is related to the electromagnetic interaction model of Fokker, 
a model where infinitesimal worldline segments have their endpoints connected by light signals. 
In this way we have elucidated not only the geometrical origin of the Doppler factor, 
but also the origin of the name, showing that the relativistic Doppler effect
is the missing link between the Doppler factor in the LW potentials and the classical Doppler effect.
Indeed, the same Minkowski diagram can be used to explain both the relativistic Doppler effect
and the Doppler factor in the LW potentials of a point particle,
the only difference being that in the first case the corresponding segments have a finite length
while in the second case they have an infinitesimal length.

\section{Acknowledgments}

The author is very much indebted 
to David~H.~Delphenich, for 
translating the articles by Fokker and Wiechert upon personal request.


\begin{thebibliography}{9}

\bibitem{lienard} Alfred-Marie~Li\'{e}nard, \lq\lq Champ \'{e}lectrique et magn\'{e}tique produit par une charge concentr\'{e}e en un point et anim\'{e}e d'un mouvement quelconque\rq\rq, {\it L'Eclairage Electrique} {\bf 16}, 5 (1898).

\bibitem{wiechert} Emil~J.~Wiechert, \lq\lq Elektrodynamishe Elementargesetze\rq\rq, {\it Archives N\'{e}erlandaises de Sciences Exactes et Naturelles} {\bf 5}, 549 (1900).

\bibitem{Griffiths4ed} David~J.~Griffiths, {\it Introduction to Electrodynamics, 4th ed.},  (Pearson, 2013), 452 and 453.

\bibitem{ORahilly} Alfred O'Rahilly, {\it Electromagnetic Theory}, (Dover, 1965), 213 and 214.

\bibitem{Aguirregabiria} J.~M.~Aguirregabiria, A.~Hern\'{a}ndez, and M.~Rivas, 
\lq\lq The Li\'{e}nard-Wiechert potential and the retarded shape of a moving sphere\rq\rq,
{\it Am. J. Phys.} {\bf 60}, 597 (1992).

\bibitem{Morse} Philip~M.~Morse and Herman~Feshbach, {\it Methods of Theoretical Physics, vol. I}, (McGraw-Hill, 1953), 214.

\bibitem{Panofsky} Wolfgang~K.~H.~Panofsky and Melba Phillips, {\it Classical Electricity and Magnetism}, 
(Addison-Wesley, 1962), 325 and 342.

\bibitem{FeynmanII} Richard~P~.Feynman, Robert~B.~Leighton, and Matthew~Sands, {\it The Feynman Lectures on Physics, vol. II}, 
(Addison-Wesley, 1964), (21-11).

\bibitem{Page} Leigh~Page and Norman~I.~Adams, {\it Electrodynamics}, (D. Van Nostrand, 1940), 149.

\bibitem{Dikshit} Biswaranjan Dikshit, \lq\lq Space–time diagram approach in derivation of Li\'{e}nard–Wiechert
potential for a moving point charge\rq\rq, {\it Can. J. Phys.} {\bf 91}, 519 (2013).

\bibitem{JoelFranklin} Joel Franklin, {\it Classical Field Theory}, (Cambridge, 2017), 48.

\bibitem{Petkov} Vesselin~Petkov, {\it Relativity and the Nature of Spacetime}, (Springer, 2005), 233.

\bibitem{Landau4ed} L.~D.~Landau and E.~M.~Lifshitz, {\it The Classical Theory of Fields, 4th ed.}, (Pergamon, 1994), 163.

\bibitem{fokker} A.~D.~Fokker, \lq\lq Ein invarianter Variationssatz f\" ur die Bewegung mehrerer elektrischer Massenteilchen.\rq\rq, 
{\it Z. Phys.} {\bf 58}, 386 (1929).

\bibitem{Jackson} J.~D.~Jackson, {\it Classical Electrodynamics, 2nd ed.}, (Wiley, 1975),   574.

\bibitem{galeriuAAAD} C\u alin~Galeriu, 
\lq\lq Time-Symmetric Action-at-a-Distance Electrodynamics and the Structure of Space-Time\rq\rq, 
{\it Physics Essays} {\bf 13}, 597 (2000).

\bibitem{galeriuArXiv} C\u alin~Galeriu, 
\lq\lq Electric charge in hyperbolic motion: arcane geometrical aspects\rq\rq, 
arXiv:1712.02213 [physics.gen-ph] (2017).

\bibitem{costa1} Olivier Costa de Beauregard, 
\lq\lq Dynamique relativiste des n points et statique classique des n fils\rq\rq,
{\it Comptes rendus de l\rq Acad\'{e}mie des Sciences} {\bf 237}, 1395 (1953).

\bibitem{minkowski2} Hermann~Minkowski, \lq\lq Die Grundgleichungen f\"ur die elektromagnetischen Vorg\"ange in bewegten K\"orpern\rq\rq, 
{\it G\"ottinger Nachrichten}, 53 (1908).

\bibitem{ScottWalter} Scott Walter, \lq\lq Breaking in the 4-vectors: the four-dimensional movement
in gravitation, 1905–1910\rq\rq, published in 
J\"{u}rgen Renn and Matthias Schemmel, editors, {\it The Genesis of General Relativity, Vol. 3}, (Springer, 2007), 193.

\bibitem{NicholasWheeler} Nicholas Wheeler, {\it Principles of Classical Electrodynamics}, (Reed College, 2002), 356. 

\bibitem{brown} Kevin~Brown, {\it Physics in Space and Time}, (Lulu, 2015), 228.

\bibitem{Neipp} C.~Neipp {\it et al.}, \lq\lq An analysis of the classical Doppler effect\rq\rq, {\it Eur. J. Phys.} {\bf 24}, 497 (2003).

\bibitem{Synge1963} J.~L.~Synge, \lq\lq Group Motions in Space-time and Doppler Effects\rq\rq, {\it Nature} {\bf 198}, 679 (1963).

\bibitem{TevianDray} Tevian Dray, {\it The Geometry of Special Relativity}, (CRC Press, 2012), 44.

\bibitem{ataman} S.~Ataman, \lq\lq Einstein's time dilation and the relativistic Doppler shift: avoiding the pitfalls\rq\rq, 
{\it Eur. J. Phys.} {\bf 42}, 025601 (2021).

\end{thebibliography}
\end{document}